\documentstyle[twoside,fleqn,espcrc2,epsf]{article}

    \newcommand{\ba}{\begin{eqnarray}}
    \newcommand{\ea}{\end{eqnarray}}
    \newcommand{\be}{\begin{equation}}
    \newcommand{\ee}{\end{equation}}

    \newcommand{\phd}{\phi^{\dagger}}
\newcommand{\AmS}{{\protect\the\textfont2
  A\kern-.1667em\lower.5ex\hbox{M}\kern-.125emS}}

\title{Study of Liapunov Exponents and the 
Reversibility of Molecular Dynamics Algorithms}

\author{
        Karl Jansen$^{\rm a}$
          and 
        Chuan Liu \address{Deutsches Elektronen Synchrotron, DESY\\          
        Notkestrasse 85, D-22603 Hamburg, Germany}%
        \thanks{Speaker at the conference.}
         }
       
\begin{document}

\begin{abstract}
We study the question of lack of reversibility and the chaotic
nature of the equations of motion in numerical simulations of lattice QCD.
\end{abstract}

\maketitle

\section{Introduction}

Molecular dynamics algorithms like
the Hybrid Monte Carlo \cite{tony} and the Kramers equation \cite{horo,kra_us}
 algorithm have been playing a major role in numerical 
simulations of QCD on a Euclidean lattice.
In order for these  algorithms to fulfill the
detailed balance condition, 
the classical motion  governed
by the set of Hamilton's equations for the system should be reversible.
However, as noticed some time ago \cite{bob}, this reversibility
condition is violated due to the round-off errors in
the numerical integration of the equations of motion.
Recently, it was pointed out \cite{kra_us}
that, due to the chaotic nature of the equations of motion, the round-off
errors in the integration which violate the
reversibility condition get magnified exponentially with a
positive Liapunov exponent $\nu$.

We will focus our discussion on
the equations that arise in the simulations of Wilson QCD
with two flavors of quarks with degenerate masses.
Periodic boundary conditions have been taken
for all the fields. The full partition function for Wilson QCD is given by,
\be
{\cal Z}
 =\int {\cal D}U{\cal D}\phd{\cal D}\phi
       \exp\;\left(- S_g-\phd Q^{-2}[U]\phi \right) \;\;,
\ee
with $S_g$ being the Wilson plaquette action.
The fermion matrix $Q[U]$ is a hermitian sparse matrix.

In molecular dynamics algorithms \cite{tony,sugar}, one introduces 
the Hamiltonian: 
\be
\label{eq:ham}
{\cal H} = \sum_{x,\mu}{1 \over 2}Tr(H^2_{x,\mu})
+ S_{eff}(U_{x,\mu},\phd,\phi)\;\;,
\ee
where $H_{x,\mu}$ is the Gaussian distributed
 momentum conjugate to the
gauge field $U_{x,\mu}$ and
takes values in $su(3)$, the Lie algebra of $SU(3)$.
We have also introduced the effective action 
$S_{eff}=S_g+\phd Q^{-2}\phi$.
Then the gauge fields and its
corresponding momenta are updated according to Hamilton's
equations of motion:
\be
\label{eq:update_continuum}
 {\dot U}_{x,\mu} =iH_{x,\mu}U_{x,\mu}\;\;, \;
i {\dot H}_{x,\mu}= [U_{x,\mu}F_{x,\mu}]_{T.A.} \;,
\ee
where the symbol $[\cdots]_{T.A.}$ stands for taking the traceless
antihermitian part of the matrix \cite{sugar} and
the quantity $U_{x,\mu}F_{x,\mu}$ is the total force
associated with the link $U_{x,\mu}$.
The dot on a field variable indicates the derivative
with respect to Monte Carlo time.
Eq.~(\ref{eq:update_continuum}) defines a Hamilton
flow in a phase space manifold which is a direct product of
$4L^3T$ factors of $SU(3)$ and $su(3)$.

\section{Liapunov Exponents}

For any flow described by a set of
first-order autonomous differential equations, the concept of
Liapunov exponents \cite{benettin1,llbook} could be introduced locally at
each point in the phase space manifold.
They describe the mean exponential
rate of divergence of two nearby trajectories.

One is thus led to study the time evolution of tangent vectors.
We take another differential of eq.~(\ref{eq:update_continuum})
and denote $dH_{x,\mu}$ and $dX_{x,\mu}=-iU^{-1}_{x,\mu}dU_{x,\mu}$
as the corresponding tangent vectors, we have:
\ba
\label{eq:update_d}
 {\dot {dX}}_{x,\mu} &=&U^{-1}_{x,\mu}dH_{x,\mu}U_{x,\mu}\;\;,
\\ \nonumber
i {\dot {dH}}_{x,\mu}&=& d[U_{x,\mu}F_{x,\mu}]_{T.A.} \;\;.
\ea
We introduce the norm in tangent space as:
\be
D^2(\tau)
= \sum_{x,\mu} tr(dH^2_{x,\mu}(\tau)+dX^2_{x,\mu}(\tau))\;\;.
\ee
The Liapunov exponent can be defined as:
\be
\nu= \lim_{D(0) \rightarrow 0}\lim_{\tau \rightarrow \infty}
{1 \over \tau}\log{ D(\tau) \over D(0) }\;\;.
\ee
The exponent with the largest real part is called the leading Liapunov exponent.

The numerical calculation of Liapunov
exponents of a given flow can be
done straightforwardly \cite{benettin1}.
Starting at $\tau=0$, we have some initial value of
$D^2(0)=\sum_{x,\mu}tr(dH^2_{x,\mu}+dX^2_{x,\mu})$ for
a given initial tangent vector $(dH_{x,\mu},dX_{x,\mu})$.
 We then integrate one step in time
with step size $\delta\tau$ using the leapfrog scheme.
Now, we evaluate the norm
$D(\delta\tau)$ of the new tangent vector and store this information.
 Next, we rescale the new tangent vector such that
its norm is still equal to $D(0)$. Repeating this for $N_{md}$ steps, we
get a sequence of values for the norm:
$D(0),D(\delta \tau),\cdots D(N_{md}\tau)$.
It can be shown \cite{benettin1} that the average
\be
\nu_n={1 \over n\delta\tau} \sum^{n}_{k=1} \log{D(k\delta\tau) \over D(0)}
\ee
is approaching the leading Liapunov exponent when $n \rightarrow \infty$.
Its value is independent of the value of $\delta\tau$, 
as long as $\delta\tau$ is not too large.

\section{Liapunov Exponents for Simulations of QCD}

\begin{figure}[t]
\vspace{-0mm}
\centerline{ \epsfysize=7.0cm
             \epsfxsize=7.0cm
             \epsfbox{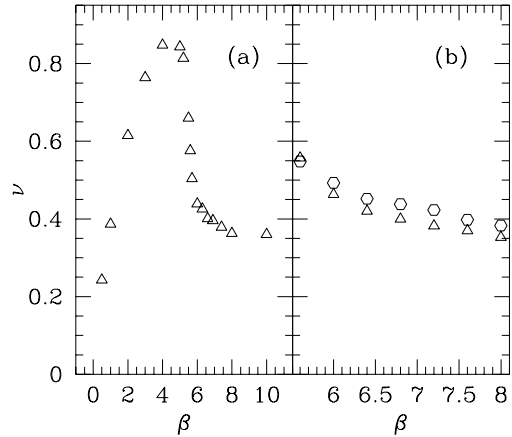}}
\vspace{-10mm}
\begin{center}
\parbox{7.5cm}{\caption{ \label{fig:expg}
The average leading Liapunov exponents for pure $SU(3)$ gauge
theory (a) and full QCD (b) as a function of
 the gauge coupling $\beta$.
In (b), exponents for different values of $\kappa$ are represented by
different symbols. The triangles correspond to $\kappa=0.13$ and
the hexagons correspond to $\kappa=0.16$.
}}
\end{center}
\end{figure}

We have studied the leading Liapunov exponents for various
values of $\beta$ and $\kappa$ on $4^4$ lattices.
In Fig.~\ref{fig:expg} (a), we plot the exponents as a function
of $\beta$ for the pure $SU(3)$ gauge theory.
\footnote{Similar studies have also been done by
the authors of ref. \cite{ivan}. Complete consistent values
of the exponent up to $\beta=10$ have been obtained. Note that
there is a factor of $\sqrt{2}$ difference between their normalization
of Monte Carlo time and ours.}
We see from this figure that
the Liapunov exponents show a significant $\beta$-dependence.
It has been conjectured recently \cite{ivan} that this
dependence might be  related to the correlation length of the theory.
However, this hypothesis is subject to further investigation.

We have also performed similar investigations of
the Liapunov exponents for full QCD with dynamical fermions in the
parameter range $5.6 \leq \beta \leq 8.0$ and $0.13 \leq \kappa \leq 1.6$.
In Fig.~\ref{fig:expg}(b), we show the final
result of the exponents as a function of $\beta$ and $\kappa$.
 For each value
of $\beta$, we have taken $4$ different values of $\kappa$, namely
$0.13$, $0.14$, $0.15$ and $0.16$. Due to their weak dependence
on $\kappa$, we only show the exponents
 for two values of $\kappa$, i.e. $\kappa=0.13$ (triangles)
and $\kappa=0.16$ (hexagons).
 The errors of the points are
about the size of the symbols.
In the parameter range that we have studied, the Liapunov exponent
is roughly $0.4-0.5$.

\section{Consequences of Irreversibility}

Due to rounding errors in the numerical integration,
the reversibility condition that is exploited in the proof of the
detailed balance condition for molecular dynamics algorithms
is violated \cite{bob,kra_us,wuppertal,brower}.
When the trajectory length $\tau$ is increasing,
we expect the strength of this violation to grow
for two reasons. The first one is due to the accumulation
of rounding errors, which is expected to grow like $\sqrt{\tau}$. 
The second cause is due to the chaotic nature of the equations
of motion which amplifies the reversibility violation like $\exp(\nu\tau)$
with a positive Liapunov exponent $\nu$.

In order to quantify the effect of these violations
in a real simulation, we have run two versions 
of HMC simultaneously
for Wilson QCD with gauge group $SU(2)$ on various
lattice volumes $V$. One of them is with 32-bit
arithmetic but with 64-bit arithmetic scalar products and summations over
the lattice, which mimics the typical situation in simulations on
a 32-bit machine. The other one runs with complete 64-bit arithmetic and
serves as a reference point for an ``exact'' program.

The Metropolis step at the end of the trajectory in molecular
dynamics algorithms depends on the value of $\Delta H$, the difference
of the Hamiltonian after and
before the integration of eq.~(\ref{eq:update_continuum}).
We have compared the two measurements of $\Delta H$, one obtained from
the  32-bit arithmetic version and the other from the 64-bit version
of the program. 
\footnote{Other physical observables 
could also be measured and compared and
we will report this in the near future.} 
The absolute value of the difference between the two
values of $\Delta H$ 
is measured after each molecular dynamics step
($<|\delta(\Delta H)_{step}|>$), and at the end
of the trajectory ($<|\delta(\Delta H)_{traj}|>$),
with the finding that $<|\delta(\Delta H)_{traj}|>$ to be
larger than $<|\delta(\Delta H)_{step}|>$.
We also observe that the value of 
$<|\delta(\Delta H)_{step}|>$
is increasing linearly with $\sqrt{V}=L^2$ while the value
of $<|\delta(\Delta H)_{traj}|>$ seems to grow faster.
On $8^4$,$12^4$ and $16^4$ lattices,
we find 
$<|\delta(\Delta H)_{traj}|>$
to be $0.08\%$, $0.18\%$ and $0.7\%$ of
$<|\Delta H|>$ respectively.  
If this trend is still maintained, 
at about a $32^4$ size lattice, 
$<|\delta(\Delta H)_{traj}|>$ could reach 
about $5\%$ of $<|\Delta H|>$ for long trajectories in 
HMC algorithm, which we think is already dangerous.
However, it seems that the quantity $<|\delta(\Delta H)_{step}|>$ will
remain reasonably small, at about 
$1\%$ of $\Delta H$ even on a $32^4$ lattice.
This suggests that running the Kramers equation algorithm will still
be feasible on such lattices with 32-bit arithmetic.

\section{Conclusions}

We have investigated several questions related to
the reversibility problem of the molecular dynamics algorithms
for the simulation of lattice QCD.
We have determined the leading
Liapunov exponents for both pure $SU(3)$ gauge theory and full
QCD for various bare parameters.
We estimated that the effects of rounding errors would become dangerous
when running the HMC algorithm on a large lattice of $O(32^4)$
 with only 32-bit precision.
In contrast, the comparably performing Kramers equation algorithm
has substantially reduced rounding error effects and its use
on such lattices is safer.

\section{Acknowledgements}

We thank
 members of the SESAM collaboration for
useful discussions.
Critical comments by  I.~Horv\'{a}th and A.~D.~Kennedy  are
gratefully acknowledged.

\end{document}